\documentclass[11pt]{article}
\usepackage{graphicx}
\usepackage{newpasp}

\begin{document}

%\vskip -1in {\em \noindent To appear in \underbar{Brown Dwarfs} (IAU 211
%Proceedings)\\
%ASP Conf.\ Series, 2002, in press}

\title{Properties of Circum(sub)stellar Accretion Disks}

\author{Michael C.\ Liu\altaffilmark{1}, Alan T.\ Tokunaga}
\affil{Institute for Astronomy, 2680 Woodlawn Drive, University of
Hawai\`{}i, Honolulu, HI 96822} \altaffiltext{1}{Beatrice Watson Parrent
Fellow. Email: mliu@ifa.hawaii.edu}

\author{Joan Najita} 
\affil{National Optical Astronomy Observatory, 950 North Cherry Avenue,
Tucson, AZ 85719}

%\index{*IC 348}
%\index{Brown dwarfs}
%\index{Planetary-mass objects}
%\index{IR spectral classification}
%\index{Q index}
%\index{Young clusters}

\begin{abstract}
%Enter your abstract here. Please limit your Abstract so that it appears
%in full on the first page.
We have completed a systematic survey for disks around young brown
dwarfs and very low mass stars.  By choosing a well-defined sample and
by obtaining sensitive thermal IR observations, we can make an unbiased
measurement of the disk fraction around such low mass objects.  We find
that $\approx$75\% of our sample show intrinsic IR excesses, indicative
of circum(sub)stellar disks.  We discuss the physical properties of
these disks and their relation to the much better studied disks around
solar-mass stars.  The high incidence of disks around substellar objects
also raises the possibility of planetary systems around brown dwarfs.
\end{abstract}

\section{A Survey for Brown Dwarf Disks}

The existence of circumstellar disks and their role in planet formation
are well established for young solar-type stars.  However, little is
known about disks around young substellar objects.  Such
circum(sub)stellar disks might provide a laboratory for studying the
physical processes of disks over a wide range of (central object and
disk) mass.  In addition, the presence of disks around young substellar
objects may be an important clue to the origin of brown dwarfs.

We have recently completed a large thermal IR ($L^\prime$-band;
3.8~\micron) survey to study the frequency and properties of disks
around young brown dwarfs and very low mass stars (Liu et~al.\ 2002; see
also Liu~2002).  Our sample comprises young ($\sim$1--3~Myr) objects in
nearby star-forming regions which have been spectroscopically classified
to be very cool, corresponding to masses of $\sim$15 to
$\sim$100~$M_{\rm Jup}$ based on current models.  As described in our
paper, the objects constitute a well-defined sample and are largely free
of selection biases.

A priori, brown dwarf disks are expected to be harder to detect than
disks around stars because of lower contrast.  Substellar objects are
less luminous and have shallower gravitational potentials. Hence, their
disks should be relatively cool and may have negligible excesses in the
commonly used $JHK$ (1.1$-$2.4~\micron) bands.  In fact, we find that
thermal IR data are {\em required} to detect most disks around young
brown dwarfs.
%$L^\prime$-band emission arises from warm dust within a few stellar
%radii ($<$0.1~AU).  
Our survey is also sensitive enough to detect brown dwarf photospheres
--- hence the absence of a disk can be discerned, and the disk frequency
of young brown dwarfs can be measured for the first time.

% - - - - - - - - - - - - - - - - - - - - - - - - - - - - %

\section{Disk Frequency and Properties}

\begin{figure}[t]
\vskip -0.1in
\centerline{\includegraphics[angle=90, width=4in]{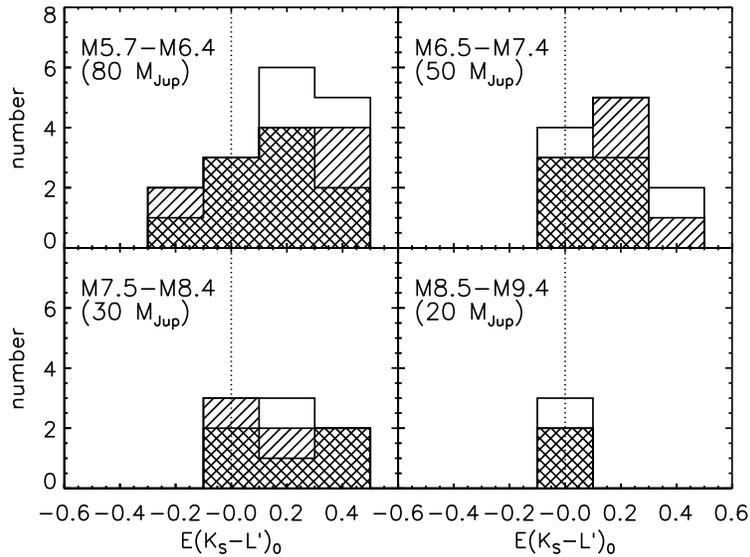}}
%\vskip 1ex
\caption{\small Histogram of intrinsic IR excesses, as measured by
dereddened $K_S\!-\!L^{\prime}$ colors, as a function of spectral
type. Approximate mass estimates are also given. The median $\pm1\sigma$
measurement errors are comparable to the bin width.  Different hatchings
represent objects from different regions. IR excesses are very common
among young brown dwarfs and low mass stars.}
\end{figure}

We find $\approx$75\% of our sample have intrinsic IR excesses: disks
around young brown dwarfs and very low mass stars appear to be common.
This disk fraction is similar to that for T~Tauri stars in the same
star-forming regions.  Figure~1 presents histograms of the IR excesses
for our sample as a function of spectral type.  The three earliest
spectral type bins (M5.7$-$M6.4, M6.5$-$M7.4, and M7.5$-$M8.4) are all
very similar, based on the K-S test.  Since the evolutionary tracks at
fixed mass are roughly constant in $T_{eff}$ for young ages, spectral
types provide a relative mass scale.  Hence, we find that the excesses
are largely independent of central mass.  The exceptions are the coolest
(lowest mass) objects, types M8.5$-$M9.4, where the excesses are small,
consistent with non-existent.

Figure~2 examines the age dependence of the disks.  If we separate
objects by spectral subclass, the absolute $K_S$-band magnitude can be
used as a surrogate for age,
%due to the fact that evolutionary tracks at
%fixed mass roughly follow constant temperature in the HR diagram.
due to the fact that model isochrones are roughly horizontal in the HR
diagram.  We find no statistically significant correlation between the
IR excesses and $M_{K_S}$ --- this suggests that the inner regions of
young brown dwarf disks do not evolve substantially over the first
$\sim$3~Myr.  This timescale is in accord with studies of disks around
T~Tauri stars (e.g. Strom et~al.\ 1989).

\begin{figure}[h]
%\hspace{0.5cm}
\centerline{\includegraphics[angle=90, width=4in]{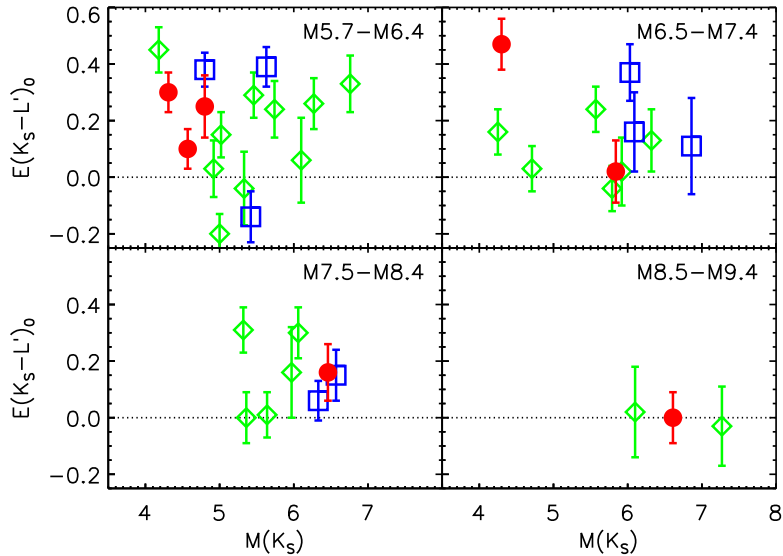}}
%\vskip -3ex
\caption{\small IR excess as a function of dereddened absolute
$K_S$-band magnitude, which is a proxy for the relative age at fixed
spectral type.  No statistically significant correlation with $M_{K_S}$
exists for each spectral type bin, suggesting the inner disk regions do
not evolve substantially over the first $\sim$3~Myr. The symbols
represent objects from different regions.  Mass estimates for each bin
are given in Figure 1.}
\end{figure}

The $L^\prime$-band flux arises from the inner disk regions ($<$0.1~AU)
and hence is sensitive to the presence of inner holes.  The disk
emission from a specific object depends on the viewing angle to the
observer, which is of course unknown.  This limitation can be overcome
with a large unbiased sample of objects --- meaningful constraints on
the inner holes can then be found from (1)~the maximum observed IR
excess and (2) the observed distribution of IR excesses.  An example of
the latter is shown in Figure~3, which compares the observed IR excess
distribution with passive (reprocessing) disk models having different
inner hole sizes.  Inner holes of $\approx 2 R_*$ are favored by the
observations; disks without inner holes would produce many more objects
with large IR excesses than are actually observed.

%To summarize, we find the observed maximum amplitude and color
%distribution of the \KmLp\ excesses point to disks with characteristic
%inner radii of $\approx2 R_*$.  

%Note that our analysis based on passive disks provides a {\em lower
%estimate} of the hole size: larger holes could be accommodated if the
%luminosity from accretion is substantial or if the vertical flaring of
%the disks is unexpectedly large.  

What is the origin of these holes?  For T~Tauri stars, inner holes are
believed to be created through the truncation of the disk by strong,
closed stellar magnetic fields; disk matter then reaches the star by
accretion along the field lines. A similar situation may be relevant to
young brown dwarfs.  Rough estimates based on magnetospheric accretion
models say that fields of a few to several hundred Gauss are needed to
produce the inner holes (see Liu et~al.\ 2002).  These values are modest
compared to those measured for pre-main-sequence stars.

%Disks with inner holes of a few stellar radii have been inferred to
%exist around the higher mass T~Tauri stars and are believed to originate
%from the interaction of the stellar magnetic field with the accretion
%disk.

%Furthermore, it is interesting that the temperature at an inner disk
%radius of $2R_*$ in our disk models is approximately 1200~K, very
%similar to the estimated temperature at the (much larger) inner edges of
%T~Tauri stars (e.g. Shu et~al.\ 1994).  At these temperatures, thermal
%ionization (of Na and K) can maintain an ionization fraction that
%enables magnetic coupling between the star (or brown dwarf) and the
%disk.  Therefore, the inner regions of brown dwarf disks will be
%sufficiently ionized for magnetic fields to couple to and truncate the
%disk at $\approx2 R_*$.

\begin{figure}[t]
\hskip -0.2in
\centerline{
\includegraphics[width=2.1in,angle=0]{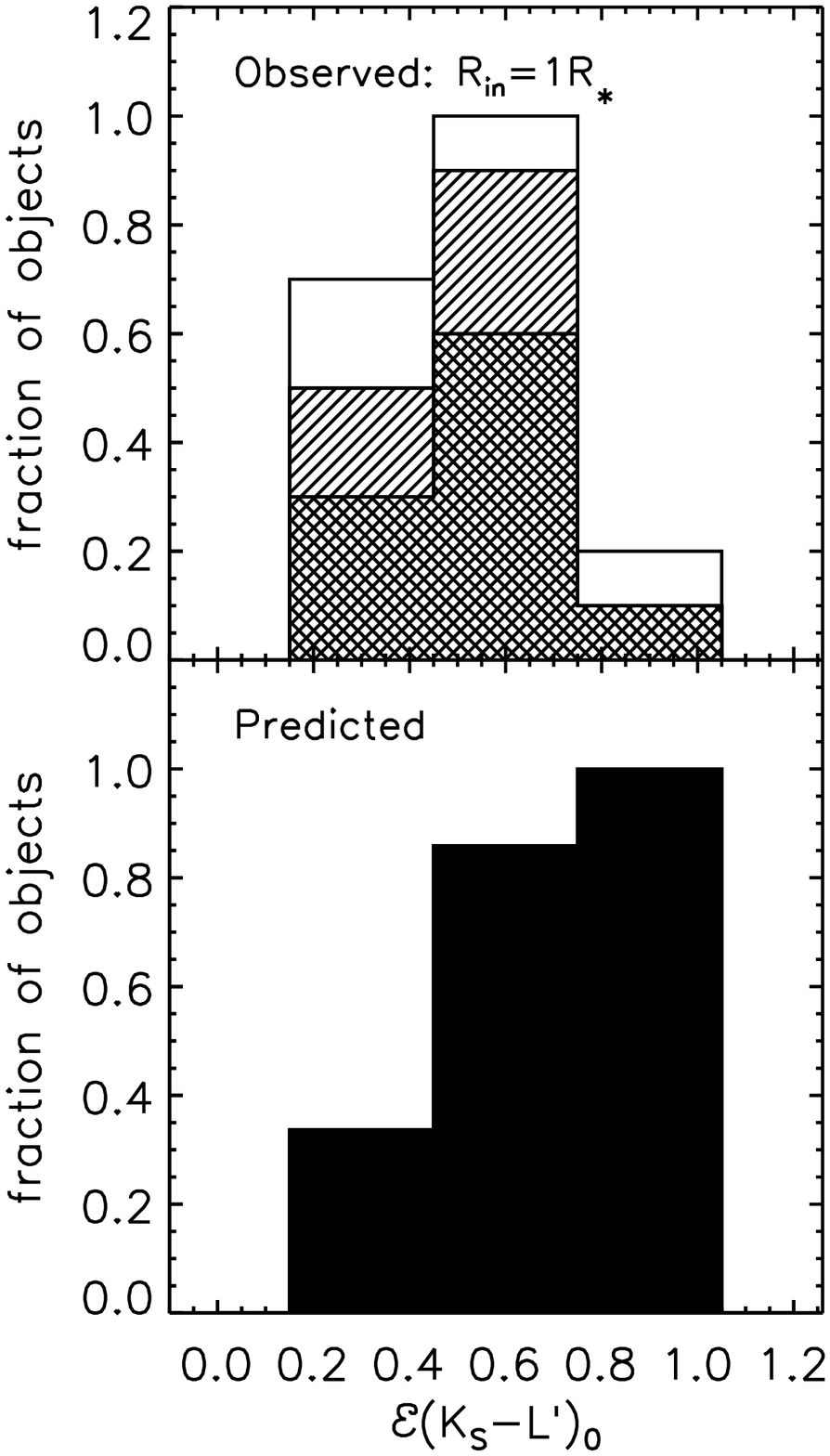}
\hskip -0.1in
\includegraphics[width=2.1in,angle=0]{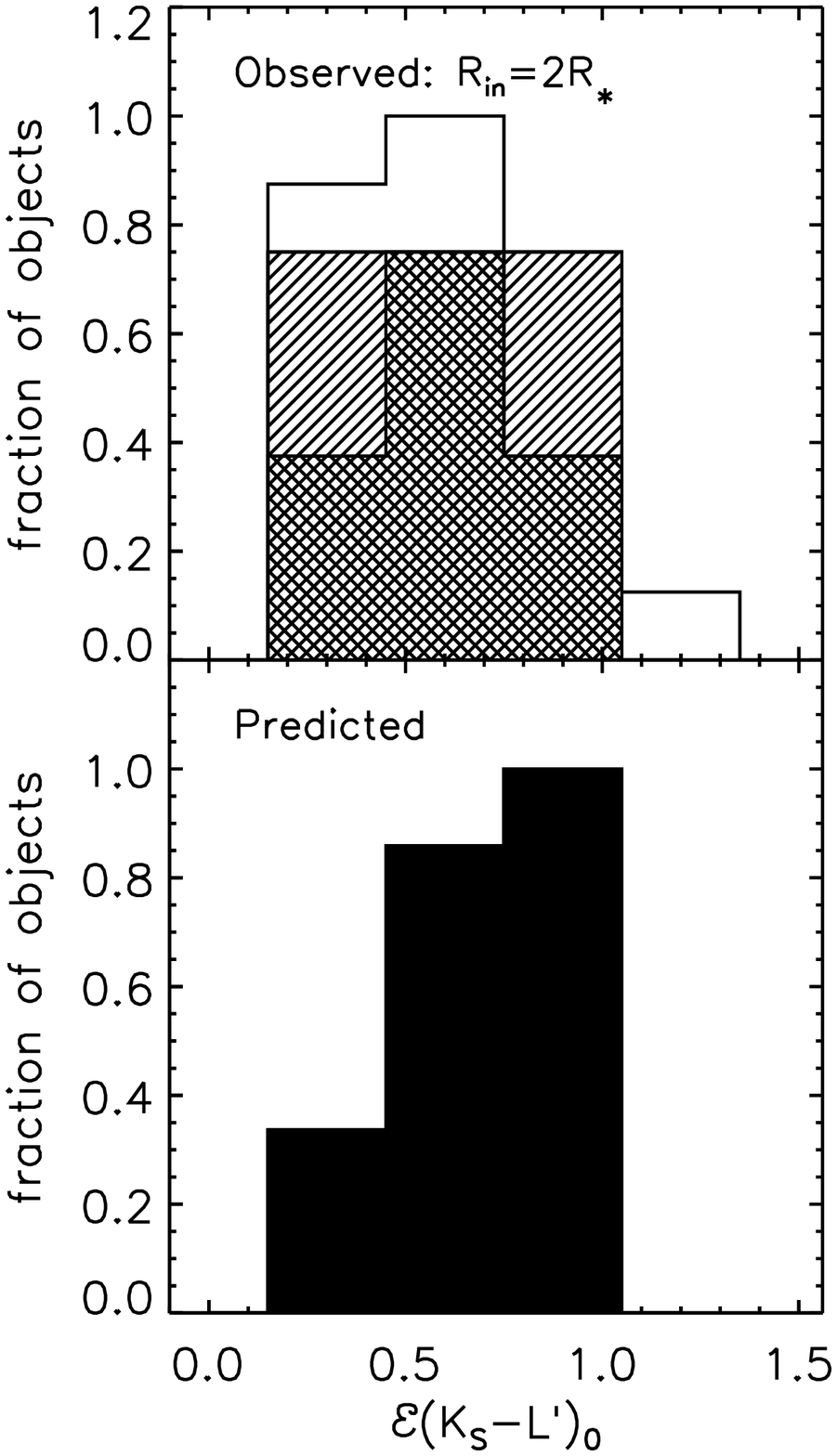}
\hskip -0.4in
}
%\vskip 2ex
\caption{\small Comparison of the observed IR excess distribution with
passive disk models possessing different inner radii.
%For a given disk model, the observed $K_S\!-\!L^{\prime}$ excess have
%been normalized by the maximum possible model excess (i.e., a face-on
%disk), forming the normalized color excess ${\cal
%E}(K_S\!-\!L^{\prime})$.
The top panels show the observations; the shadings indicate different
sub-samples.  The bottom panels show the predicted excess distribution
for disk models seen from randomly chosen viewing angles.  Models
without inner holes ({\em left plot}) predict a much larger fraction of
IR excesses than observed.  The observations agree better with models
with an inner radius of $\approx2 R_*$ ({\em right plot}), suggesting
that brown dwarf disks have modest inner holes.  See Liu et~al.\ (2002)
for details.}
\end{figure}

\section{Connections to Disks around Young Solar-Type Stars}

It appears that much of the observational paradigm developed for T~Tauri
disks can be extended to disks around young brown dwarfs.  Based on data
from the literature, much of our sample shows H$\alpha$ emission,
suggesting that most young brown dwarfs are accreting, although the
inferred rates are much lower than for T~Tauri stars (e.g. Muzerolle
et~al. 2000).  Most of our objects show both IR excesses and strong
H$\alpha$ emission, like the higher mass classical T~Tauri stars. Others
have little/no IR excess and little H$\alpha$ emission, analogous to
weak-line T~Tauri stars.
%A few have IR excesses but only weak H$\alpha$ emission; such a
%phenomenon is also seen among weak-line T~Tauri stars
%\citep[e.g.][]{1989AJ.....97.1451S, 1990AJ.....99.1187S,
%2002AJ....123.1613P}.  

We also find that brown dwarf disks, like those of T~Tauri stars,
possess inner holes.  While the hole sizes are very different for
T~Tauri stars and brown dwarfs, the estimated disk {\em temperatures} at
the inner radii are similar, of order 1200~K.  Moreover, the inner disks
are hot enough for significant thermal ionization (of Na and K), which
is needed to permit coupling of the magnetic field to the disk.
% and are hot enough for significant thermal
%ionization, which is needed for coupling of the magnetic field to the
%disk.  
Thus, our inferred hole sizes suggest that the magnetic accretion
paradigm developed for T~Tauri stars may extend to substellar masses.

%Thus, our inferred hole sizes suggest that the magnetic accretion
%paradigm developed for T~Tauri stars may extend to substellar masses.
%This illustrates one way in which brown dwarf disks may be useful
%laboratories for testing our understanding of the physical processes of
%circumstellar disks.

Finally, the high disk fraction of young brown dwarfs raises the
possibility of forming planets around these objects.  Such planetary
systems would represent a fascinating alternative to the numerous
systems found around solar-type stars.

%\acknowledgments We thank the SOC and LOC for organizing a very
%enjoyable and stimulating conference. This work has been supported by
%the Beatrice Watson Parrent Foundation.


\begin{references}
\parskip=1pt
%\small
%\reference Haisch, K., Lada, E., \& Lada, C. 2001, AJ, 121, 2065
%\reference Kenyon, S. \& Hartmann, L. 1995, ApJS, 101, 117
%\reference Muench, A., Alves, J., Lada, C. \& Lada, E. 2001, ApJ, 558, L51
\reference Liu, M. 2002, in {\em Proc.\ of IAU 211: Brown Dwarfs},
in press (astro-ph/0207477)

\reference Liu, M., Najita, J. \& Tokunaga, A. 2002, ApJ, submitted

\reference Muzerolle, J. et al.\ 2000, ApJ, 545, L141 

\reference Strom, K., Strom, S., Edwards, S., Cabrit, S. \& Struksie,
M. 1989, AJ, 97, 1451

\end{references}
\end{document}